\documentclass[11pt]{article}
\usepackage[margin=2.54cm]{geometry}

\date{}

\usepackage{amsmath,amsthm, color, tikz}
\usepackage{tcolorbox}
\usepackage{thmtools} 
\usepackage{bbm}
\usepackage{dsfont}
\usepackage{balance}
\usetikzlibrary{positioning}
\usetikzlibrary{shapes}
\usetikzlibrary{decorations.pathreplacing}
\usepackage{enumerate, enumitem} 
\usepackage{hyperref}

\usepackage{mathtools}

\usepackage{latexsym}
\usepackage{graphicx}
\usepackage{amsmath}
\usepackage{fullpage}
\usepackage{tikz}
\usepackage[ruled, lined, linesnumbered, commentsnumbered, longend, ]{algorithm2e}
\usepackage{mathrsfs}  
\usepackage{xcolor}

\newtheorem{theorem}{Theorem}[section]
\newtheorem{fact}[theorem]{Fact}

\newtheorem{lemma}[theorem]{Lemma}
\newtheorem{definition}[theorem]{Definition}

\newtheorem{remark}[theorem]{Remark}

\usepackage{algpseudocode}

\hypersetup{colorlinks,linkcolor={blue},citecolor={blue},urlcolor={red}}

\usepackage{authblk}
\usepackage{bbm} %
\usepackage{mathtools}

\usepackage[colorinlistoftodos,prependcaption,textsize=tiny]{todonotes}

\newcommand{\pr}{\mathsf{Pr}}

\newcommand{\ex}{\mathbf{Ex}}
\newcommand{\set}{\textsc{set}}
\newcommand{\score}{\textsc{scr}}

\newcommand{\algo}{\mathscr{A}}

\newcommand{\ep}{\mathsf{EP}}
\newcommand{\error}{\textsc{Error}}

\newcommand{\don}[1]{\mathrm{Don#1}}

\newcommand{\cvmb}{\mathrm{CVM2}}
\newcommand{\cvmap}{\mathrm{CVM1'}}
\newcommand{\cvmbp}{\mathrm{CVM2'}}

\newcommand{\widesim}[2][1.5]{
  \mathrel{\overset{#2}{\scalebox{#1}[1]{$\sim$}}}
 }
\newcommand{\iid}{\widesim[2]{i.i.d}}

\newcommand{\final}{\textsc{Last}}

\title{Analysis of Knuth's Sampling Algorithm D and D'}

\author[1,2]{Mridul Nandi}
\author[1]{Soumit Pal}
\affil[1]{Indian Statistical Insititue, Kolkata}
\affil[2]{IAI, TCG-CREST}
	
\date{\today}

\begin{document}

\maketitle

\begin{abstract}
In this research paper, we address the Distinct Elements estimation problem in the context of streaming algorithms. The problem involves estimating the number of distinct elements in a given data stream $\mathcal{A} = (a_1, a_2,\ldots, a_m)$, where $a_i \in \{1, 2, \ldots, n\}$. Over the past four decades, the Distinct Elements problem has received considerable attention, theoretically and empirically, leading to the development of space-optimal algorithms. A recent sampling-based algorithm proposed by Chakraborty et al.~\cite{chakraborty_et_al:LIPIcs.ESA.2022.34} has garnered significant interest and has even attracted the attention of renowned computer scientist Donald E. Knuth, who wrote an article on the same topic~\cite{don:knuth:CVM} and called the algorithm CVM.
In this paper, we thoroughly examine the algorithms (referred to as CVM1, CVM2 in~\cite{chakraborty_et_al:LIPIcs.ESA.2022.34} and DonD, $\don{D'}$ in~\cite{don:knuth:CVM}). 
We first unify all these algorithms and call them cutoff-based algorithms. Then we provide an approximation and biasedness analysis of these algorithms. 


\end{abstract}




\section{Introduction}
\noindent\textsc{Notations}. We adopt the notation $x^t := (x_1, \ldots, x_t)$ for any $t$-tuple, whenever there is no ambiguity in the context (i.e., it can be clearly distinguished from a power of a number or a set). For a positive integer $n$, let $ [n] := \{1,2, \ldots, n\}$ be a universe. We simply write the statement $(1-\varepsilon) c \leq x \leq (1 + \varepsilon) c$ as $x = (1 \pm \varepsilon)c $ (or $x \neq (1 \pm \varepsilon)c$ to denote the negation).  \medskip



\paragraph{Distinct Elements Problem (or DEP) for data streams.}  Given a data stream $a^m := (a_1, a_2, \ldots, a_m)$ of $ m$ elements from the universe, we need to output an estimate $\hat{F_0}$ of $ F_0 := F_0(a^m) := |\{a_1, \ldots, a_m \}|$. Throughout this paper we fix an arbitrary stream $a^m$ for some positive integer $m$.

\subsection{Goals of Estimates}
We are interested in developing streaming algorithms for all these problems mentioned above with low space complexities.
We consider two qualitative natures of approximations of the estimates. 
 \smallskip

\begin{itemize}
\item 
\noindent $(\epsilon, \delta )$-approximation: An estimate $\hat{F_0}$ of $F_0$ is called an $(\epsilon, \delta)$-approximation if 
$$\pr[\hat{F_0} \neq (1 \pm \epsilon) \cdot F_0] \leq \delta.$$ 

\item 
\noindent $\delta$-biased estimation: 
We call the estimate $\hat{F_0}$  $\delta$-biased $|\ex(\hat{F_0}) - F_0| \leq \delta$. If $\delta = 0$, we call it {\em unbiased} estimation. 
\end{itemize}

\subsection{Cutoff-based Algorithm}
Consider the following scenario: Candidates appear in a queue for an interview, where each candidate can appear multiple times. The selection board can choose a maximum of $s$ candidates at any given time and so it maintains a list of selected candidates so far. The $t$th candidate, denoted as $a_t$, receives a score $q_t \in [0,1]$. The selection board sets a cutoff, denoted as $p_t$, based on the current score $q_t$, the previous cutoff $p_{t-1}$, and the list $L_{t-1}$ just before candidate $a_t$ appears. The list may hold some additional information such as score for each selected candidate in the list. In our context, obtaining a low score in the final appearance is the eligibility criterion for being selected in the list. So, if $q_t \geq p_t$, the candidate $a_t$ is rejected, regardless of whether they were in the previous list or not. Furthermore, this selection process (we call it filter operation) will be applied to all existing candidates in the list $L_{t-1}$ (i.e., candidates with scores greater than or equal to $p_t$ are removed from the list). 

-- We have mentioned that the cutoff is chosen adaptively (we call it cutoff update function). However, the cutoff must be chosen in such a way that at each time $t$, the list $L_t$ does not exceed $s$ candidates. 

-- It is assumed that candidates have a forgetful nature, meaning their score in each interview appearance is independent of their previous scores.  We model a distribution (called score distribution) representing the distribution of the scores for all candidates. \smallskip

This process continues until all candidates $a_1, a_2, \ldots, a_m$ have been processed, resulting in the final list $L_m$ and final cutoff $p_m$. Consequently, the score obtained in the final interview holds the most significance for selection, as previous scores are not considered. This last score is referred to as the "credit score" for a candidate. However, it is important to note that previous non-credit scores also play a role, as they influence the values of the cutoff and indirectly impact the selection process. We formalize this process in algorithmic language and refer to it as the cutoff-based streaming algorithm.

The cutoff-based streaming algorithm serves as a simplified abstraction of recent algorithms proposed by Don Knuth, namely Algorithm D (referred to as DonD) and Algorithm D$'$ (referred to as $\don{D'}$)~\cite{don:knuth:CVM}. It is worth noting that the CVM algorithms from~\cite{chakraborty_et_al:LIPIcs.ESA.2022.34} can also be viewed as equivalent representations within the same framework. 
In this framework, a list refers to a set of the form $\{(x_1, r_1), \ldots, (x_t, r_t)\}$, where the $x_i$'s are distinct elements of $A$, and $r_i \in [0,1]$.
To describe the cutoff-based streaming algorithm, we introduce the following operations for a list $L$, $p \in [0,1]$:
\begin{itemize}
\item $\textsc{Remove}(L, a) = \{(a', q') \in L: a' \neq a\}$ (remove $a$ if it is there)

\item $\textsc{Filter}(L, p) = \{(a, q) \in L: q < p\}$ (a subset of $L$ with score less than the given cutoff)

\item $\set(L) = \{a \in [n]: \exists q, (a, q) \in L\}$  (set of elements ignoring the scores)

\item $\max_{\score}(L, p) = \max(L) = \max\{q: (a, q) \in L\}$. \medskip
\end{itemize}

\  \smallskip

\noindent\underline{\textbf{Algorithm} $\textsc{Cutoff}[D, C]$}. \smallskip

\noindent It uses a distribution $D$, called {\em score distribution},  over the sample space $[0,1]$, and a cutoff update function $C$ (which takes a list $L$ and a cutoff value as input and returns a new cutoff value). This also uses a {\em bucket limit parameter} $s$, a positive integer to be decided by the algorithm (the list size should not cross the bucket limit). It initializes $L_0 = \emptyset$ and $p_0 = 1$. Upon receiving an element $a_t$ at time $t \in [m]$, the algorithm follows the following steps:


\begin{enumerate}
\item[step-1] (sample score of $a_t$): $q_t \sim D$. 

\item[step-2] (do we include $a_t$?): 
$$L'_t = \begin{cases}
    \textsc{Remove}(L_{t-1}, a_t) & \textit { if } q_t \geq p_{t-1} \\
    \textsc{Remove}(L_{t-1}, a_t) \cup (a_t, q_t) & \textit{ otherwise}.
\end{cases} 
$$

\item[step-3] (we are done if we have space): If $|L'_{t}| \leq s$ then \\ {\bf return} $p_t = p_{t-1}$ and $L_t = L'_{t}$. \smallskip

\item[step-4] (if not, then resample the list to reduce to our limit): Else (i.e., $|L'_t| > s$)  {\bf return}
$$p_t := C(L'_{t}, p_{t-1})\footnote{Here we assume that the pair $(a_t, q_t)$ is marked in the last $L'_t$ so that the cutoff update function can use it explicitly. The $p_t$ should be chosen in a way so that $L_t$ must reduce with probability almost one.},~~~ L_t = \textsc{Filter}(L'_{t}, p_t).$$ 



\end{enumerate}
\noindent After processing all elements, a final estimate is 0 if $L_m = \emptyset$ or $\hat{F_0} := |L_m|/ D([0,p_m))$, whenever $D([0,p_m)) > 0$ (otherwise, we return $n$ or $m$ whichever is smaller and known to the algorithm).

\begin{definition}
	Let $\algo := \textsc{Cutoff}[D, C]$ for some $D$ and $C$. We note that the behavior of the algorithm is completely determined by $q^m$ (scores of all appearances). So we write $\algo(q^m) = (L_t, p_t)_{ t \in [m]}$ and we call it transcript of the process. 
\end{definition}

\subsection{Our Contribution}
In this paper, we provide an analysis of the unbiasedness of all these algorithms and also an approximation analysis based on some intriguing observations on the DonD algorithm. 



\section{Some Known Algorithms}

\subsection{Score Distributions}
A distribution $D$  over $[0,1]$ is called linear if there is a subset $\Omega \subseteq [0,1]$ such that for all $x \in \Omega$, $D([0, x) )  = x$ and $D(\Omega) = 1$.  For example, uniform distribution $U$ over $[0,1]$ is linear with $\Omega = [0,1]$.  We will see an example of discrete linear distribution.  \smallskip

\noindent\textsc{Uniform and Discrete Uniform}. For every $N > 0$, let $U_N$ denote the discrete version of uniform distribution over the set 
$$\mathcal{U}_{N} := \{0, \frac{1}{N}, \ldots, \frac{N-1}{N} \},$$ for some large $N$.We write $U$ to denote the continuous uniform distribution over $[0,1]$. It is well known that $U_N$ converges to $U$ (in distribution).  \medskip

\noindent\textsc{Geometric and Geometric like Distributions}. 
Let $0< p < 1$. We write $X \sim Ber(p)$ (Bernoulli distribution) if $\pr(X=1) = p = 1 - \pr(X=0)$. We write $X \sim Geo(p)$ (Geometric distribution) if for all $k \geq 1$, $\pr(X = k) = (1-p)^{k-1}p$. Let $X_1, X_2, \ldots \iid Ber(p)$ be a process of Bernoulli trials. Let $Y$ denote the first $k$ for which $X_k = 1$, then $Y \sim Geo(p)$. 

Let $N'$ be a fixed positive integer. A {\em Truncated Geometric} random variable $X \sim TGeo(p, N')$ satisfies the following: $\pr(X = k) := (1-p)^{k-1}p$ for all $k \in [N']$ and $\pr(X = N'+1)  :=  (1- p)^{N'}$. This is same running Bernoulli process with probability $p$ and stop once we obtain 1 or the number of trials becomes $N'+1$.. In this setup, the number of trials follows the truncated Geometric distribution. 

\begin{definition}[Geometric like distributions]
Probability distribution $G_{N'}$ and $G_{\infty}$ over the support
\[ \mathcal{G}_{N'}  := \{2^{-N'-1}, 2^{-N'}, \ldots, 2^{-2}, 2^{-1} \}, ~~ \mathcal{G}_{\infty}  := \{2^{-k} : k \geq 1 \} \]
(closely related to the truncation geometric distributions with a truncation at $N' > 0$ and the geometric distribution) are described respectively, as follows: 
\[G_{N'}(2^{-k}) = 
\begin{cases}
    2^{-k}  & \textit{ if } \forall k \in [N'] \\
    2^{-N'} & \textit{ if }  k = N'+1
\end{cases}
\]
\[G_{\infty}(2^{-k}) = 2^{-k}, ~~ \forall k \geq 1.\]
\end{definition}

We define functions $g_{N'}  : [0,1] \to \mathcal{G}_{N'}$ and $g_{\infty}  : [0,1] \to \mathcal{G}_{\infty}$
as follows: 
$$ g_{N'}(x)= \begin{cases}
	2^{-k} & \textit{ if } 2^{-k} \leq x < 2^{-k+1}, k \in [N'] \\
	2^{-N'-1} & \textit{ if }  0 \leq x \leq 2^{-N'-1}
\end{cases}
$$
\[ g_{\infty}(x)  = 2^{-k} ~~ \textit{ if } 2^{-k} \leq x < 2^{-k+1} \]

Let $N = 2^{N'}$. Note that for all $x \neq 0$, 
\begin{equation}
\frac{x}{2} \leq  g_{N'}(x) \leq x
\end{equation}
and if $X \sim U_{2^{N'}}$, $Y \sim U$ then $g_{N'}(X) \sim G_{N'}$ and $g_{\infty}(Y) \sim G_{\infty}$. Note that $G_{\infty}$ is also linear. 


\subsection{$\don{D}$ Algorithm}
DonD is same as $\textsc{Cutoff}[U, \max_{\score}]$. When we actually implement the continuous distribution, we must have to use a distribution modeled as $U_N$ for some large $N$, usually a power of 2.  We write $\don{D}_{\mathrm{disc}}$ to denote the discrete version of DonD (the value of $N$ would be implicitly understood), namely $\textsc{Cutoff}[U_N, \max_{\score}]$.  We first observe the following for any execution of the algorithm:
For all $t$,  $|L_t| \leq s$ and the else-condition (i.e., step-4) will be applied only if $|L'_t| = s+1$, $a_t \in \set(L'_t)$ and $a_t \not\in \set(L_{t-1})$.  The whole statement can be proved by induction on $t$.

\subsection{Other Cutoff-based Algorithms}

\begin{enumerate}
	\item $\don{D'}$ is same as $\textsc{Cutoff}[G_{\infty}, \max_{\score}]$. Note that the original description of $\don{D'}$ is $\textsc{Cutoff}[U, \max_{\score, g}]$ where $\max_{\score, g}(L, p) = \max(g(L))$. However, it is not difficult to verify that for all $u^m$, we have $(\set(L_t),  p_t) = (\set(M_t), p'_t)$ for all $t$ where $(L_t, p_t)_t$ and $(M_t, p'_t)_t$ are the transcript of these two variants with score tuple $u^m$ and $q^m$ respectively where $q_i = g(u_i)$.  Moreover, note that 
	$$u_1, \ldots, u_m \iid U_{2^{N'}} ~~\Rightarrow ~~~ q_1, \ldots, q_m  \iid G_{\infty}.$$ 
	We write $\don{D'}_{\mathrm{disc}} := \textsc{Cutoff}[G_{N'}, \max_{\score}]$ to denote the discrete version of $\don{D'}$ (the value of $N'$ would be implicitly understood).

	\item CVM1 (Algorithm 1 of~\cite{chakraborty_et_al:LIPIcs.ESA.2022.34}) is essentially same as $\cvmap := \textsc{Cutoff}[G_{\infty}, C_{\cvmap}]$ where
	$$C_{\cvmap}(L'_{t}, p_{t-1}) = \begin{cases}
		p_{t-1}/2 & \textit{ if } \max_{\score}(L'_{t}) = p_{t-1}/2\\
		\mathrm{Abort} & \textit{ otherwise.}  
	\end{cases}
	$$
	Once abort, the algorithm terminates and can return an arbitrary estimates. \smallskip
	
	\item CVM2 (Algorithm 2 of~\cite{chakraborty_et_al:LIPIcs.ESA.2022.34}) is essentially same as $\cvmbp := \textsc{Cutoff}[G_{\infty}, C_{\cvmbp}]$ where
	$$p_t := C_{\cvmbp}(L'_{t}, p_{t-1}) =  p_{t-1}/2.$$
\end{enumerate}


\section{Analysis of Estimates}\label{sec:estimate}

\subsection{Fair Algorithm}
\noindent\textsc{Notations}. For any $t \in [m]$ and $j \in [t]$, we denote $\final(a_j, t)$ as the maximum value of $k \in [t]$ for which $a_k = a_j$. Additionally, we define the set 
$$\mathcal{F}_t = { \final(a_1, t), \final(a_2, t), \ldots, \final(a_t, t)}$$ to capture all the time points up to $t$ at which the elements appear for the last time (i.e., they do not appear again until time $t$). We simply write $\mathcal{F} = \mathcal{F}_m$ for brevity. 

For a subset $F \subseteq [m]$, we write $q^F$, a subtuple of $q^m$, to denote the tuple $(q_i: i \in F)$. For any set $S \subseteq A$, we write $\mathcal{F}|_S := \{i \in \mathcal{F}: a_i \in S\}$. For all $t \in \mathcal{F}$, we call $q_t$ credit score. However, all non-credit scores may also have some influence on determining the final cutoff $p_m$. 
For any integer $r \in [F_0]$,  
$c_r := q_{(r)}^{\mathcal{F}}$ is the $r$-th order element of $q^{\mathcal{F}}$ (i.e., the $r$th smallest credit scores among all credit scores). We write $T_r$ to denote the time point at which $c_r$ score appears and hence $q_{T_r} = c_r$. \medskip

If $F_0 \leq s$ then $p_m = 1$ as we do not have to execute the step-4.  In this case the estimate would be perfect. So we assume $F_0 > s$ and hence $p < 1$. Moreover, $p = q_j$ for some $j$.  


\begin{definition}[fair cutoff sampling algorithm]\label{defn:fair}
	A cutoff-based sampling algorithm $\algo$ is said to be {\em fair} if  $\forall t \in [m], j \in \mathcal{F}_t$
	\[ a_j \in L_t \Leftrightarrow  q_j < p_t.\]
\end{definition}

\begin{lemma}[Lemma M of~\cite{don:knuth:CVM}]
	Let $\algo$ be a cutoff-based algorithm such that cutoff values are decreasing (i.e., non-increasing). Then, $\algo$ is a fair algorithm.  
\end{lemma}
The proof of the lemma is more or less straightforward from the definition of the cutoff-based algorithm. Lemma M of~\cite{don:knuth:CVM} says that DonD is a fair algorithm. However, we have shown any decreasing cutoff-based algorithm, in particular DonD, is a fair algorithm. It is easy to verify that CVM1, CVM2, $\don{D}$ and $\don{D'}$ algorithms have decreasing cutoff functions and hence 
\begin{center}
	CVM1, CVM2, and $\don{D}$ and $\don{D'}$ are fair algorithms. 
\end{center}

\begin{lemma}\label{don-prop1}
Let $s$ be the bucket limit parameter for $\don{D}$ and $S \subseteq A$ be a set of size $x \leq s$. There is a function $J: [0,1]^{m-x} \to [m]$ satisfying the following:.
\[\forall q^{[m] \setminus \mathcal{F}|_S} = \alpha, ~~~ \big( S \subseteq \set(L_m)  \big) ~~\Leftrightarrow~~ \big(p_m = q_{J(\alpha)}  ~~\wedge~~ q_t < q_{J(\alpha)} ~\forall t \in \mathcal{F}|_S\big). \] 
\end{lemma}

\proof Let $s_j = q_i$ for all $j \not\in \mathcal{F}|_S$ and $s_{j} = 0$, otherwise. We write $(M_t, p'_t)_t$ to denote the transcript of the modified process of DonD where $s^m$ is used as the score tuple.  We claim that for all $t \in [m]$, (i) $\set(L_t) = \set(M_t)$ and $p_t = p'_t$.  This can be proved by induction on $t$. The base case is obvious and let us assume the statement for $t-1$. Now, we have two cases: $t \in \mathcal{F}|_S$. In this case, we must include $a_t$ in $L_t$ (as it is the last chance to be included for $a_t$). Clearly, we include $a_t$ in $M_t$. Now, if we execute step-4 then it has to be applied for both execution and the maximum values (which are now $p_t$ and $p'_t$) do not depend on $q_t$ and $s_t$. So, $p'_t = p_t$. 
So there is some $j$, for which $q_j = p'_m$ and the value of $j$ depends only on $s^m$ (and so depends only on $\alpha$ and the set $S$ itself).  This proves the one direction of the statement. The other direction is obvious as $q_t < q_{J(\alpha)} = p_m$. \qed \medskip

Later we state a more stronger statement. 

\begin{remark}
Note that the above statement is true for any cutoff-based algorithms of the form  $\textsc{Cutoff}[D, \max_{\score}]$ and $\textsc{Cutoff}[D, C_{\cvmb}]$  for any score distribution $D$ (as we see it is not used in the proof).
\end{remark}

\subsection{Unbiased Estimator}

\begin{theorem}\label{thm:unbiased}
		Algorithm $\don{D}$, $\don{D'}$ and $\cvmb$ return an unbiased estimate.		
%
\end{theorem}

\proof Fix $a \in A$ and let $J$ be the function as defined in the above Lemma~\ref{don-prop1} with $S = \{a\}$. Let $i = \final(a, m)$ and  $I_a$ be the indicator random variable to denote the event that $a \in \set(L_m)$. Now, for any $\alpha$, conditioned on $ q^{[m]\setminus i} = \alpha$, $I_a = 1$ if and only if $p_m = q_j$ and $q_i < q_j$ where $j = J(\alpha)$. So the conditional random variable $I_a/p_m$ can take only nonzero value $1/q_j$ with probability $\pr(q_i < q_j~|~ q^{[m]\setminus i} = \alpha)$ which is $q_i$ (as $D$ is linear). Hence, $\ex(I_a/p_m) = 1$ and so $\ex(|L_m|/p_m) = |A|$. \qed \medskip

\section{$(\epsilon, \delta)$-Approximation Analysis}

\subsection{Basic Setup}

\subsubsection{Chernoff Bound}
We recall a fact (one form of {\em Chernoff bound}) which would be used in the analysis. 
\begin{fact}[Chernoff Bound (Theorem 4.4(2), 4.5(2) of~\cite{10.5555/1076315})]\label{fact-chernoff}
	Let $N$ be a positive integer and $ 0 < p, \epsilon \leq 1$. Then,
	\begin{align*}
		\ep^-(N, p, \epsilon) := \sum_{x = 0}^{Np(1 - \epsilon)} \binom{N}{x} p^x(1-p)^{N-x} & \leq  e^{-Np \epsilon^2/2} \\
		\ep^+(N, p, \epsilon) := \sum_{x = Np(1 + \epsilon)}^{N} \binom{N}{x} p^x(1-p)^{N-x} & \leq  e^{- Np \epsilon^2/3}.
	\end{align*}
\end{fact}

So for any $N, p, \epsilon, \beta > 0$, 
\begin{equation}\label{eq:ep1}
	Np \geq 3 \epsilon^{-2} \log \beta^{-1} \Rightarrow \ep^+(N, p, \epsilon), \ep^-(N, p, \epsilon) \leq \beta.
\end{equation}

\subsection{Error Event Analysis: For small $p_m$}
Now we state a technical lemma and using this we provide a method to analyze $(\epsilon, \delta)$-approximation. The lemma follows from the observation that $q_1, q_2, \ldots q_m$ are i.i.d. with distribution $D$.  

\begin{lemma}[technical lemma]
Let $B \subseteq A := \{a_1, \ldots, a_m\}$. Then, for any fair algorithm using the linear score distribution $D$, and $x \leq s$,
\begin{align*}
\pr[\set(L_m) = B \wedge p_m = p] & \leq p^{|B|}(1-p)^{|A|- |B|} \\
\pr[|\set(L_m)| = x \wedge p_m = p] & \leq {F_0 \choose x} p^{|B|}(1-p)^{|A|- |B|} 
\end{align*}
\end{lemma}


\begin{definition}[Error Events]
For a real number $ p \in [0,1]$, we define the event  $E_{p}^{+}$ to denote the following event: $$\frac{|\set(L_m)|}{p_m} > F_0(1 + \epsilon) ~~~\wedge~~~ p_m =p.$$ 
Similarly we define the event $ E_{p}^{-} $ to denote the event:
$$\frac{|\set(L_m)|}{p_m} < F_0(1 - \epsilon) ~~~\wedge~~~ p_m = p.$$
We further denote 
$$E^+_{\geq p_0} = \bigcup_{\substack{p \geq p_0 \\ p \in \Omega}} E^+_{p},~~ := E^-_{\geq p_0} = \bigcup_{\substack{p \geq p_0 \\ p \in \Omega}} E^-_{p}.$$
We write the union event $E_{\geq p_0} := E^+_{\geq p_0} \vee E^-_{\geq p_0}$. So the error event $\error$ is the union of the event $E_{\geq p_0}$ and $(p_m < p_0)$ for any suitable choice of $p_0$. 
\end{definition}

\begin{lemma}\label{lem:cvm2:prob}
    Let $s \geq 12 \log\frac{2m}{\delta}$, $k = \lceil \log_2 (2F_0/s) \rceil$ and $p_0 := 2^{-k} > 0$ (so $s/2 \geq F_0 p_0 \geq s/4$). For $\cvmbp$ algorithm, we obtain $$\pr(p_m<p_0) \leq \delta/2.$$ 
\end{lemma}

\begin{proof}
Note, $p_m < p_0$ event holds only if for some $ t \in [m]$ we have $ p_t < 2^{-k}$ for the first time, i.e., $p_{t-1}= 2^{-k} \land p_t= 2^{-k-1}$. 
and hence we must have 
$|\set(L_{t-1})| \geq s$. So, we have the following equations:
\begin{align*}
\pr(p_{t-1}= 2^{-k} \land p_{t} = 2^{-k-1}) & \leq \pr(|\set(L_{t-1})| \geq s \land p_{t-1}= p_0) \\
& \leq \sum_{x\geq s} \binom{F_0}{x} p_0^{x} (1- p_0)^{F_0-x}\\
& \leq \ep^+(F_0, p_0, 1) \leq \delta/2m ~~~~\textit{as } F_0p_0 \geq 3 \log\frac{2m}{\delta}
\end{align*}
So by using the union bound over all values of $k$, the result follows. \medskip
\end{proof}

If we choose $s \geq  12 \log\frac{4m}{\delta}$ then we have $\pr(p_m < p_0) \leq \delta/4$. For the same choice of $s$, it is easy to see that $\cvmap$ aborts with probability at most $m 2^{-s} \leq \delta/4$. Note that if it does not abort then $\cvmbp$ and $\don{D'}$ behave identically and hence, we can have the following lemma. 

\begin{lemma}
Let $s \geq  12 \log\frac{4m}{\delta}$, $k = \lceil \log_2 (2F_0/s) \rceil$ and $p_0 = 2^{-k}$. Then, 
$\pr(p_m < p_0)$ in $\don{D'}$ algorithm is at most $\delta/2$. Moreover, in $\cvmap$, $\pr(p_m < p_0 \vee \mathrm{Abort}) \leq \delta/2$.
\end{lemma}

\noindent\textsc{Relationship between $\don{D}$ and $\don{D'}$.}
Suppose we sample $u_1, u_2, \ldots u_m \iid U_{2^{N'}}$ and then we define $g(u_i) = q_i \iid G_{N'}$. Now we run $\don{D}$ using $u^m$ and $\don{D'}$ using $q^m$. Let $p_t$ and $p'_t$ denote the cutoff probability for DonD and $\don{D'}$ respectively. 
Then, it is easy to verify that (also stated in~\cite{don:knuth:CVM})
\[ \forall t \in [m], p'_t \leq p_t \leq 2p'_t. \]
Using this and the above lemma for $\don{D'}$, we have the following lemma. 
\begin{lemma}\label{lem:cvm2:prob}
    Let $s \geq 24 \log\frac{4m}{\delta}$, $k = \lceil \log_2 (2F_0/s) \rceil$ and $p_0 := 2^{-k} > 0$. For $\don{D}$ algorithm, we obtain $\pr(p_m<p_0) \leq \delta/2.$ 
\end{lemma}







\subsubsection{Error Event Analysis: $E_{\geq p_0}$}
We note that $\pr(\error) \leq \pr(p_m < p_0) + \pr(E_{\geq p_0})$ and hence it is sufficient to bound $\pr(E_{\geq p_0})$.

\begin{theorem}[$G$ score Distribution]\label{thm:generic-bound}
Let $\algo$ be an {\em decreasing (hence fair) cutoff} based algorithm using {\em score distribution} $G_{N'}$ (over a sample space $\Omega$) with a bucket limit parameter $s =  \max\{24\log\frac{4m}{\delta}~,~ \frac{6}{\epsilon^2} \log\frac{8}{\delta}\}$, 
and  $p_0 = 2^{-\lceil \log_2 (2F_0/s) \rceil}$ (so $F_0 p_0 \geq \frac{3}{\epsilon^2} \log\frac{8}{\delta}\}$).  
Then, 
\[ \pr(E_{\geq p_0}) \leq \delta/2.\]
Hence, for the same choice of $s$, $\don{D'}, \cvmap, \cvmbp$ are $(\epsilon, \delta)$-approximation algorithms. 

\end{theorem}

\proof For every $p = 2^{i} p_0$, $i \geq 0$, $\pr(E^+_{p}), \pr(E^-_{p}) \leq (\delta/8)^{2^i}$ (follows from the above technical lemma and the variant of Chernoff bound, Fact~\ref{fact-chernoff}). The result follows by summing over all terms (bounded by the geometric sum).\qed \medskip

Analysis of bounding $\pr(E_{\geq p_0})$ for DonD requires a different technique as we do not have a geometric sum (recall, DonD uses uniform distribution). We postpone the analysis for later.




\section{Approximation Analysis for $\don{D}$}
\begin{theorem}
Algorithm $\don{D}$ is an $(\epsilon, \delta)$-estimator for 
$$s =  \max\{ 24\log\frac{4m}{\delta},~~  \frac{24}{\epsilon^2} \log \frac{96}{\epsilon^2 \delta} \} $$
\end{theorem}

We have already seen that for all $s \geq 24 \log\frac{2m}{\delta}$, $k = \lceil \log_2 (2F_0/s) \rceil$ and $p_0 := 2^{-k} > 0$
$$\pr(p_m<p_0) \leq \delta/2.$$
The rest of the section will be devoted to bound $\pr(E_\geq p_0)$ for a suitable choice of $s$.

\subsection{Dependency of Final Cutoff}
We now further extend the idea used in the previous section by fixing only $q^{[m] \setminus \mathcal{F}}$ values instead of $q^{[m]\setminus j}$. 
We still use {\em maximum function as a cutoff update function $C$} (as DonD). Let $S$ be a fixed set of size $x \leq s$. Now, given any $q^m$, we define a modified tuple of scores $s^m$ as follows: 
$$s_i = \begin{cases}
q_i & \textit{ if } i \not\in \mathcal{F}\\
0 & \textit{ if } i \in \mathcal{F}, a_i \in S \\
1 & \textit{ if } i \in \mathcal{F}, a_i \not\in S 
\end{cases}
$$
To modify the credits of each element, we assign a credit of 0 if the element belongs to set $S$, and a credit of 1 otherwise. Non-credit scores remain unchanged. We denote the tuples of lists and cutoffs using the revised scores as $(M_t, p'_t)_{t \in [m]} := \algo(s^m) := \algo'(q^{[m] \setminus \mathcal{F}})$. We are interested in exploring the relationship between the original transcript $(L_t, p_t)_{t \in [m]}$ and the transcript of the revised process $(M_t, p'_t)_{t \in [m]}$ under the condition that $\set(L_m) = S$.

In the original scores, the credits of elements in $S$ are strictly lower than the credits of elements in $S^c$. Let $(c_a)_{a \in A}$ denote the tuple of credits (for the original scores, noting that the credits for the revised scores are either 0 or 1). There exists an ordering of elements in $A$, denoted as $(a'_1, a'_2, \ldots, a'_{F_0})$, such that $$c_1 := c_{a'_1} < c_2 := c_{a'_2} < \cdots < c_{F_0} := c_{a'_{F_0}}, ~~ S = \{ a'_1, \ldots, a'_{x}\},~~ c_{x} < p_m \leq c_{x+1}.$$ We denote the time at which $a'_{x+1}$ appears for the last time as $t_0$. Hence, $q_{t_0} = c_{x+1}$.

The first change can occur only at time $t \in \mathcal{F}|_{S^c}$ where $a_t \in S^c$ is clearly rejected in $M_t$, but $L_t$ may still include $a_t$. This implies that the value of $p_t$ may be smaller or equal to $p'_t$. By induction on $t$, we can prove the following simple claim. \medskip



\noindent\textbf{Claim}. For all $t \in [m]$, 
$p_t \leq p'_t$ and $\textsc{Filter}(M_t, p_t) = L_t \setminus \mathcal{F}|_{S^c}$.  \medskip

Let $t_1$ denote the first time $t$ such that $p_t \leq q_{t_0}$ (and hence onward the elements of $S^c$ appearing for the last time will be rejected in the original transcript).
Using this observation, we clearly have two possibilities:

\begin{enumerate}
    \item  $p_{t_1-1} > c_{x+1}$ and $p_{t_1} < c_{x+1}$. In this case we have $\textsc{Filter}(M_{t_1}, p_{t_1}) = L_{t_1}$ and no elements in $(a, q) \in M_t$ can have $q > p_{t_1}$ (as we have applied the step-4 in the original algorithm at time $t_1$). So, $p'_{t_1} = p_{t_1}$. Hence onward these two cutoff values will remain equal to each other and so $p_m = p'_m < c_{x+1}$.

\item $p_{t_1-1} > c_{x+1}$ and $p_{t_1} = c_{x+1}$. Now we can have again two subcases:

\begin{enumerate}
    \item At time $t_2 > t_1$, for the first time, $p_{t_2} < c_{x+1}$. Once again, we can apply same argument as above and we have  $p_m = p'_m < c_{x+1}$. 
    
    \item $p_t = c_{x+1}$ for all $t \geq t_1$. In this case $p_m = c_{x+1} <  p'_m$. Note that $p'_m \neq p_m$ as the revise algorithm has credit 1 instead of $c_{x+1}$. 
\end{enumerate}
\end{enumerate}

We now summarize the above discussion as follows. 

\begin{theorem}
We fix a set $S$ of size $x \leq s$. We sample any $q^{\mathcal{F}}$ such that all $q$ values for elements of $S$ are less than for all $q$ values for $S^c$. Let $p'_m$ be the final cutoff in the execution of $\algo'(q^{[m]\setminus\mathcal{F}})$ (the revised algorithm substituting credits of elements by 0 or 1). Let $p_m$ denote the original cutoff value. Then, the following hold:

-- If $c_{x+1} \geq  p'_m$ if and only if $p_m = p'_m \leq c_{x+1}$.

-- If $c_{x+1} <  p'_m$ if and only if $p_m = c_{x+1} < p'_m$.
\end{theorem}

\subsection{Probability Bound of Error Event $E_{\geq p_0}^+$}


Let $N' = N(1 + \epsilon)$. Now, $E_{\geq p_0}^+$ means that $p_0 \leq p_m \leq x/N'$ and so $x \geq N'p_0 \geq s/4$.

For a set $S$ of size $|S| =x \geq s$, $q^{[m]\setminus \mathcal{F}}$, we have defined $p'_m$. Now, 
consider two conditional events (we have seen that one of these two events must occur to have $S$ as the final set):
\begin{enumerate}
    \item[C1:]  $q^{[m]\setminus \mathcal{F}}$ and $c_{x+1} \geq p'_m$.
    \item[C2:]  $q^{[m]\setminus \mathcal{F}}$, $t_0$, $q_{t_0} = c_{x+1} = p$ for a fixed $p < p'_m$.
\end{enumerate}

\noindent\textsc{Case $C1$:} We have $p_m = p'_m$ and hence we have the following:

    



\begin{align*}
     \pr(\set(L_m)=S \wedge E_{\geq p_0}^+ ~|~ C1) \leq
     \begin{cases}
         (p'_m)^x(1-p'_m)^{N-x} ~~~~~~~~~&\text{if}~~~ p_0 \leq p'_m \leq \frac{x}{N^{\prime}}\\
        0 ~~~~~~&\text{otherwise} 
     \end{cases}
\end{align*}

Since the function $h(p)= p^x (1-p)^{N-x}$ achieves maximum at $p=x/N$ and increasing up to $x/N$, the above probability is maximized at $p'_m = x/N'$. So,
\begin{align*}
 \pr(|\set(L_m)|=x \wedge E_{\geq p_0}^+ ~|~ C1) & \leq \binom{N}{x} (x/N')^{x} (1-x/N')^{N-x} \\
 &  \leq e^{\frac{-\epsilon^2 N (x/N')}{3}} \\
 & \leq  e^{\frac{-\epsilon^2 s}{24}} 
\end{align*}
By using simple calculus, we have the following: $$ s \geq \frac{24}{\epsilon^2} \log \frac{96}{\epsilon^2 \delta} \Rightarrow e^{\frac{-\epsilon^2 s}{24}} \leq \delta/4s.$$
So,
\[ \pr(E_{\geq p_0}^+ ~|~ C1) \leq \sum_{x \geq s/4}^s \pr(|\set(L_m)|=x \wedge E_{\geq p_0}^+ ~|~ C1) \leq \delta/4.\]

\noindent\textsc{Case $C2$:} We have $p_m = c_{x+1} = p$. Now, for all $S$ with $a_{t_0} \not\in S$ (otherwise, the probability will be zero) we have the following,
\begin{align*}
     \pr(\set(L_m)=S \wedge E_{\geq p_0}^+~|~ C2) \leq
     \begin{cases}
         p^x(1-p)^{N-1-x} ~~~~~~~~~&\text{if}~~~ p_0 \leq p \leq \frac{x}{N^{\prime}}\\
        0 ~~~~~~&\text{otherwise} 
     \end{cases}
\end{align*}
Now we can vary all sets $S$ such that $a_{t_0} \not\in S$ and hence there are ${N-1 \choose x}$ such sets of size $x$. Once again it is maximized at $p = x/N'$. So by using a similar argument we have 
\begin{equation}\label{eq:later2}
 \pr(|\set(L_m)|=x \wedge E_{\geq p_0}^+ ~|~ C2) \leq \binom{N-1}{x} (x/N')^{x} (1-x/N')^{N-x} 
\end{equation}
As $x \geq (1 + \epsilon)(N-1)x/N'$, we once again apply the same argument to conclude that  $\pr(E_{\geq p_0}^+ ~|~ C2) \leq \delta/4$ and hence 
$\pr(E_{\geq p_0}^+) \leq \delta/4$.

\subsection{Probability Bound of Error Event $E_{\geq p_0}^-$}

Now we assume $N' = N(1 - \epsilon)$. Let $S \subseteq A$ of size $x \leq F_0$. Now, $E_{\geq p_0}^-$ means that $p_{\max} := \max\{p_0, x/N'\} \leq p_m$


Now, for any $S \subseteq A$ of size $x \leq F_0$, let $ p_{\max}= \max\{ \frac{x}{N'}, p_0\}$.
We have
\begin{align*}
\pr(\set(L_m)= S \wedge E_{\geq p_0}^-~|~ C1) \leq
     \begin{cases}
         (p'_m)^x(1-p'_m)^{N-x} ~~~~~~~~~&\text{if}~~~ p_{\max} \leq p'_m \\
        0 ~~~~~~&\text{otherwise} 
     \end{cases}
\end{align*}
and 
\begin{align*}
\pr(\set(L_m)= S \wedge E_{\geq p_0}^-~|~ C2) \leq
     \begin{cases}
         p^x(1-p)^{N-1-x} ~~~~~~~~~&\text{if}~~~ p_{\max} \leq p \\
        0 ~~~~~~&\text{otherwise} 
     \end{cases}
\end{align*}

The above probabilities are maximized at
$ p_{\max}$ and  we also have $ Np_{\max} > \frac{x}{1-\epsilon}$ and $ Np_{\max} \geq Np_0 \geq s/4$. So, by using a similar argument, we have 

\begin{align*}
\pr(|\set(L_m)| = x \wedge E_{\geq p_0}^{-} ~~|~~ C1) & \leq  \binom{N}{x} (p_{\max})^{x} (1- p_{\max})^{N-x}\\
       & \leq   e^{\frac{-\epsilon^2s}{12}}  ~~~~\text{by Chernoff bound}\\
       &\leq (\delta/4s)^2.
\end{align*}
So summing over choices of $x \in \{0,1 ,2 \ldots, s\}$, we have 
\begin{align*}
\pr(E_{\geq p_0}^{-} ~~|~~ C1) &\leq \delta/4.
\end{align*}
Similarly, we have 
\begin{align*}
\pr(E_{\geq p_0}^{-} ~~|~~ C2) &\leq \delta/4.
\end{align*}
This completes the proof.

\section{Related Works}

\noindent\textsc{Hash-based Approach}.~~ The $F_0$ estimation problem has received extensive attention within the data streaming model \cite{AMS99, BJKST2002, bjkst_stoc_2002, blasiok2018, DurandFlajolet2003, 10.1145/1807085.1807094}. Although this problem finds applications across various computing domains, it was first explored in the algorithms community by Flajolet and Martin \cite{FM83}. They provided an initial approximation by assuming the existence of hash functions with complete independence with optimal space complexity of $O(\log n + \frac{1}{\epsilon^2} \log \frac{1}{\delta} )$.



The pioneering work by Alon, Matias, and Szegedy \cite{AMS99} introduced the data streaming model of computation and revisited the distinct elements problem as a specific case of $F_k$ estimation. They achieved a space complexity of $O(\log n)$ for $\epsilon > 1$ and constant $\delta$. Subsequently, Gibbson and Tirthpura (GT2001) proposed the first $(\epsilon, \delta)$ approximation algorithm for the distinct elements problem with space complexity of $O(\frac{\log n}{\epsilon^2})$. Bar-Yossef, Jayram, Kumar, Sivakumar, and Trevisan \cite{BJKST2002} further improved the space complexity to $\Tilde{O}(\log n + \frac{1}{\epsilon^2})$.\footnote{Note that $\Omega_{d}\left(\cdot\right)$ and $O_{d}\left(\cdot\right)$ hide factors that only depend on $d$.}.  Kane, Nelson, and Woodruff \cite{KNW2010} achieved the optimal space complexity of $O(\log n + \frac{1}{\epsilon^2})$ with respect to both $n$ and $\epsilon$. It should be noted that these complexity bounds assume a fixed confidence parameter $\delta$, which can be amplified by running $\log \frac{1}{\delta}$-estimators in parallel and returning the median, introducing a multiplicative factor of $\log \frac{1}{\delta}$.

In a significant contribution, Błasiok \cite{blasiok2018} presented an approximation algorithm for the $F_0$ estimation problem, achieving a space complexity of $O(\frac{1}{\epsilon^2} \cdot \log \frac{1}{\delta} + \log n)$. Notably, this result matches the lower bound for all three parameters, $n$, $\epsilon$, and $\delta$. \medskip

\noindent\textsc{Sampling-based Approach}.  A crucial technical aspect common to the previous works is the careful utilization of limited-independence hash functions to achieve a space complexity of $poly(\log n)$. While Monte Carlo-based approaches have been employed for estimating the size of set unions, their direct adaptation to the streaming setting has not yielded significant progress. The ground-breaking work by Gibbons and Tirthapura (\cite{Gibbons_Tirthapura_2001}) invented a sampling-based framework to compute the union of two sets in a distributed setup. 
Recently, a novel sampling-based approach was proposed for estimating the size of set unions in the streaming model, achieving a space complexity with a $\log m$-dependence \cite{10.1145/3452021.3458333}. 
In a subsequent work \cite{10.1145/3517804.3526222}, the authors provided an algorithm with space complexity independent of $m$ (but depends on $n$). 


\begin{thebibliography}{}
\bibitem{AMS99}
Noga Alon, Yossi Matias, and Mario Szegedy. The space complexity of approximating
	the frequency moments. Journal of Computer and System Sciences, 58(1):137–147, 1999.
	
	


\bibitem{bjkst_stoc_2002}
Ziv Bar-Yossef, Ravi Kumar, and D. Sivakumar. Reductions in streaming algorithms, with an application to counting triangles in graphs. In Proceedings of the Thirteenth Annual ACM-SIAM Symposium on Discrete Algorithms, SODA ’02, page 623–632, USA, 2002. Society for Industrial and Applied Mathematics.

\bibitem{BJKST2002}
Ziv Bar-Yossef, T. S. Jayram, Ravi Kumar, D. Sivakumar, and Luca Trevisan. Counting distinct elements in a data stream. In Jose D. P. Rolim and Salil Vadhan, editors, Randomization and Approximation Techniques in Computer Science, pages 1–10, Berlin, Heidelberg, 2002.


\bibitem{blasiok2018}
Jaroslaw Blasiok. Optimal streaming and tracking distinct elements with high probability, Proceedings of the 2018 Annual ACM-SIAM Symposium on Discrete Algorithms (SODA), pages 2432–2448. 


\bibitem{Chakraborty2023DistinctEI}
Sourav Chakraborty,Distinct Elements in Streams: An Algorithm for the (Text) Book,
Embedded Systems and Applications, 2023

\bibitem{chakraborty_et_al:LIPIcs.ESA.2022.34}
Sourav Chakraborty, N. V. Vinodchandran, and Kuldeep S. Meel. Distinct Elements in
Streams: An Algorithm for the (Text) Book. In Shiri Chechik, Gonzalo Navarro, Eva
Rotenberg, and Grzegorz Herman, editors, 30th Annual European Symposium on Algorithms (ESA 2022), volume 244 of Leibniz International Proceedings in Informatics (LIPIcs),
pages 34:1–34:6, Dagstuhl, Germany, 2022.



\bibitem{DurandFlajolet2003}
Marianne Durand and Philippe Flajolet. Loglog counting of large cardinalities. In Giuseppe
Di Battista and Uri Zwick, editors, Algorithms - ESA 2003, pages 605–617, Berlin, Heidelberg,
2003


\bibitem{FM83} 
Philippe Flajolet and G. Nigel Martin. Probabilistic counting algorithms for data base
applications. Journal of Computer and System Sciences, 31(2):182–209, 1985.




\bibitem{Gibbons_Tirthapura_2001}
Phillip B. Gibbons and Srikanta Tirthapura. Estimating simple functions on the union of data streams. Proceedings of the thirteenth annual ACM symposium on Parallel algorithms and architectures, Jul 2001.



\bibitem{10.1145/1807085.1807094}
Daniel M. Kane, Jelani Nelson, and David P. Woodruff. An optimal algorithm for the distinct elements problem. PODS ’10, page 41–52, New York, NY, USA, 2010. 

\bibitem{don:knuth:CVM}
Donald E. Knuth. The cvm algorithm for estimating distinct elements in streams. (25
May 2023, revised 26 May 2023). 


\bibitem{KNW2010} 
Daniel  Kane. Jelani Nelson and David P Woodruff,
An Optimal Algorithm for the Distinct Elements Problem,
PODS'2010, Association for Computing Machinery, 2010. 


\bibitem{10.1145/3517804.3526222}
Kuldeep S. Meel, Sourav Chakraborty, and N. V. Vinodchandran. Estimation of the size of union of delphic sets: Achieving independence from stream size. PODS’22, page 41–52, New York, NY, USA, 2022.



\bibitem{10.1145/3452021.3458333}
Kuldeep S. Meel, N.V. Vinodchandran, and Sourav Chakraborty. Estimating the size of union of sets in streaming models. PODS’21, page 126–137, New York, NY, USA, 2021. 




\bibitem{10.5555/1076315}
Michael Mitzenmacher and Eli Upfal. Probability and Computing: Randomized Algorithms
and Probabilistic Analysis. Cambridge University Press, USA, 2005.


\end{thebibliography}
\end{document}